\documentclass[12pt,a4paper]{article}
\usepackage[left=1in,right=1in,top=1.5in,bottom=1.5in]{geometry}
\usepackage{graphicx, color, setspace, amssymb, amsmath, amsthm, amstext, booktabs, fancyhdr, multirow, float, bm, lineno, times, rotating, algorithm, url, nameref} 
\usepackage[sectionbib,authoryear]{natbib}
\PassOptionsToPackage{hyphens}{url}
\usepackage[colorlinks,linkcolor=blue,citecolor=blue,urlcolor=blue]{hyperref}
\allowdisplaybreaks 
\usepackage{placeins} 
\usepackage{authblk} 
\usepackage{natbib}
\usepackage{mathtools}
\usepackage{caption}
\usepackage{times}
\usepackage{verbatim}

\newcommand{\bmat}{\begin{pmatrix}}
\newcommand{\emat}{\end{pmatrix}}

\usepackage{booktabs}
\usepackage{multirow}
\usepackage{float}

\newcommand{\bo}[1]{\boldsymbol{#1}}

\newcommand*\patchAmsMathEnvironmentForLineno[1]{%
  \expandafter\let\csname old#1\expandafter\endcsname\csname #1\endcsname
  \expandafter\let\csname oldend#1\expandafter\endcsname\csname end#1\endcsname
  \renewenvironment{#1}%
     {\linenomath\csname old#1\endcsname}%
     {\csname oldend#1\endcsname\endlinenomath}}%
\newcommand*\patchBothAmsMathEnvironmentsForLineno[1]{%
  \patchAmsMathEnvironmentForLineno{#1}%
  \patchAmsMathEnvironmentForLineno{#1*}}%

\AtBeginDocument{%
\patchBothAmsMathEnvironmentsForLineno{equation}%
\patchBothAmsMathEnvironmentsForLineno{align}
\patchBothAmsMathEnvironmentsForLineno{flalign}%
\patchBothAmsMathEnvironmentsForLineno{alignat}%
\patchBothAmsMathEnvironmentsForLineno{gather}%
\patchBothAmsMathEnvironmentsForLineno{multline}%
}

\makeatletter
\g@addto@macro\normalsize{%
  \setlength\abovedisplayskip{8pt}
  \setlength\belowdisplayskip{8pt}
  \setlength\abovedisplayshortskip{8pt}
  \setlength\belowdisplayshortskip{8pt}
}
\makeatother

\begin{document}
\title{A Comparison of Joint Species Distribution Models for Percent Cover Data}
\author[1]{Pekka Korhonen\thanks{Corresponding author: pekka.o.korhonen@jyu.fi; Department of Mathematics and Statistics, FI-40014 University of Jyv\"askyl\"a, Finland}} 
\author[2]{Francis K.C. Hui} 
\author[3,4]{Jenni Niku} 
\author[1]{Sara Taskinen}
\author[5]{Bert van der Veen} 
\affil[1]{Department of Mathematics and Statistics, University of Jyv\"askyl\"a, Finland}
\affil[2]{Research School of Finance, Actuarial Studies and Statistics, The Australian National University, Canberra, Australia}
\affil[3]{Faculty of Sport and Health Sciences, University of Jyv\"askyl\"a, Finland}
\affil[4]{Department of Biological and Environmental Science, University of Jyv\"askyl\"a, Finland}
\affil[5]{Department of Mathematical Sciences, Norwegian University of Science and Technology, Trondheim, Norway}

\date{}
\maketitle
\thispagestyle{empty}

\vspace{1cm}

\setcounter{page}{1}

\section*{Abstract}
\begin{quote}
\begin{enumerate}
\item Joint species distribution models (JSDMs) have gained considerable traction among ecologists over the past decade, due to their capacity to answer a wide range of questions at both the species- and the community-level.
The family of generalized linear latent variable models in particular has proven popular for building JSDMs, being able to    
handle many response types including presence-absence data, biomass, overdispersed and/or zero-inflated counts. 
\item We extend latent variable models to handle percent cover data, with vegetation, sessile invertebrate, and macroalgal cover data representing the prime examples of such data arising in community ecology.
\item Sparsity is a commonly encountered challenge with percent cover data. Responses are typically recorded as percentages covered per plot, though some species may be completely absent or present, i.e., have $0\%$ or $100\%$ cover respectively, rendering the use of beta distribution inadequate.
\item We propose two JSDMs suitable for percent cover data, namely a hurdle beta model and an ordered beta model. We compare the two proposed approaches to a beta distribution for shifted responses, transformed presence-absence data, and an ordinal model for percent cover classes. Results demonstrate the hurdle beta JSDM was generally the most accurate at retrieving the latent variables and predicting ecological percent cover data.
\end{enumerate}

{\emph{Keywords: beta regression, community-level modelling, latent variable model, ordination, percent cover data, zero-inflation}}

 
\end{quote}

\section*{Introduction} \label{sec:intro}

Measurements of percent cover are typical in many ecological studies of plants communities, macroalgae, or sessile animals. By their nature, e.g. limited seed dispersal, tendency for clumping, and lack of self-locomotion, the notion of 'individual' may not always be meaningful or easy to determine regarding such organisms. In such cases, is is often sensible to use percentage covered (of a given study area) by species as its measure of abundance, rather than counts or simple presence/absence. For instance, data on percent cover at a given study site are typically collected by taking measurements on multiple plots or along line transects. These  measurement can vary a lot in manner: percent cover may be determined visually by practitioners, possibly through aggregation of standardized subplots, or through pin-point methods, i.e., by placing a given amount of pins randomly across the study area and recording the proportion of 'hits' for each species part of the study \citep{damgaard2020model}. Instead of representing cover as a percentage, cover data may also (classically) be represented as ordered classes e.g., using the Braun-Blanquet \citep{braun1932plant} or Daubenmire \citep{daubenmire1959canopy} scale. Finally, with improving technologies, automated percentage cover data collection procedures based on high-resolution images (say) are expected to increase in the future. For a comprehensive review of methods to measure vegetation cover data, we refer the reader to \cite{damgaard2019using}. In this  article, we focus on the analysis of percent cover data where species are allowed to overlap each other, i.e., the sum covered by all species in a plot can exceed 100 percent.

For percent cover data, regression assuming a beta distribution offers a natural starting choice for modelling.
For cover class data, a reasonable default would be  the cumulative logit model \citep[also known as proportional odds model,][]{mccullagh80}.
On the other hand, proper analysis of cover data is often hindered by high percentages of observations recorded to be zero i.e., the responses are sparse. This causes issues particularly with models for continuous data, such as the beta or Dirichlet regression, which are unable to accommodate zero responses altogether. If the amount of zeros is relatively low, or if instead of being structural the zeros are the result of inadequate sampling, one can replace them with some small values or via imputation. However, when the zeros are structural, i.e. related to the actual underlying ecological process in question, hurdle models should instead be considered. We refer the reader to \cite{blasco2019does} and references therein for a more detailed discussion about the differences between structural zeros and sampling zeros.

In this paper, we investigate the analysis of percent cover data in the context of joint species distribution modelling \citep[JSDM,][]{pollock2014understanding, warton2015so,ovaskainen2020joint,hui23cbfm}. JSDMs are a powerful approach to analyse various types of community composition data, providing researchers with a general framework to draw inferences about e.g. co-occurrence patterns between different species, community covariation and its attribution to environmental filtering versus possible biotic interactions, and model-based ordinations on species assemblages. There is a suite of statistical software available for fitting (various flavors of) JSDMs e.g., the \textsf{R}-packages \texttt{boral} \citep{hui2018boral}, 
\texttt{HMSC} \citep{Tikhonovetal:2020}, \texttt{gllvm} \citep{Nikuetal:2019b}, and \texttt{VAST} \citep{thorson2019guidance}. 
All of these adopt generalized linear latent variable models \citep[GLLVM,][]{skrondal2004generalized} as the basis for fitting JSDMs, using either Bayesian Markov Chain Monte Carlo or approximate likelihood-based methods for estimation and inference. This article focuses on the latter, in particular the \texttt{gllvm} package \citep{gllvmpackage} which combines
variational approximations \citep[see][and references therein]{korhonen2023}
with automatic differentiation techniques from the \texttt{TMB} library \citep{tmb}
to facilitate computation efficient and scalable estimation 

Although a flexible framework by design, research and readily available implementations of JSDMs/GLLVMs for percent cover data specifically have been relatively lacking. Exceptions to this are the works of \cite{damgaard2020model,damgaard2023erratum}, who proposed a method for model-based ordination of pin-point cover data utilizing a re-parameterized Dirichlet-multinomial distribution. This is appropriate for data that, instead of percentages, includes the counts of the 'hits' of the pins. As such, their model is unable to account for structural zeros. More recently, \cite{kettunen23jsdm} introduce a similar type of model with a more general structure, letting some subsets of species to be in direct competition for space, meaning they cannot overlap one another, while simultaneously allowing it for others. In this article, we consider JSDMs for percent cover setting with a focus on model-based ordination. 

\begin{figure}[h]
    \centering
    \includegraphics[width=\textwidth]{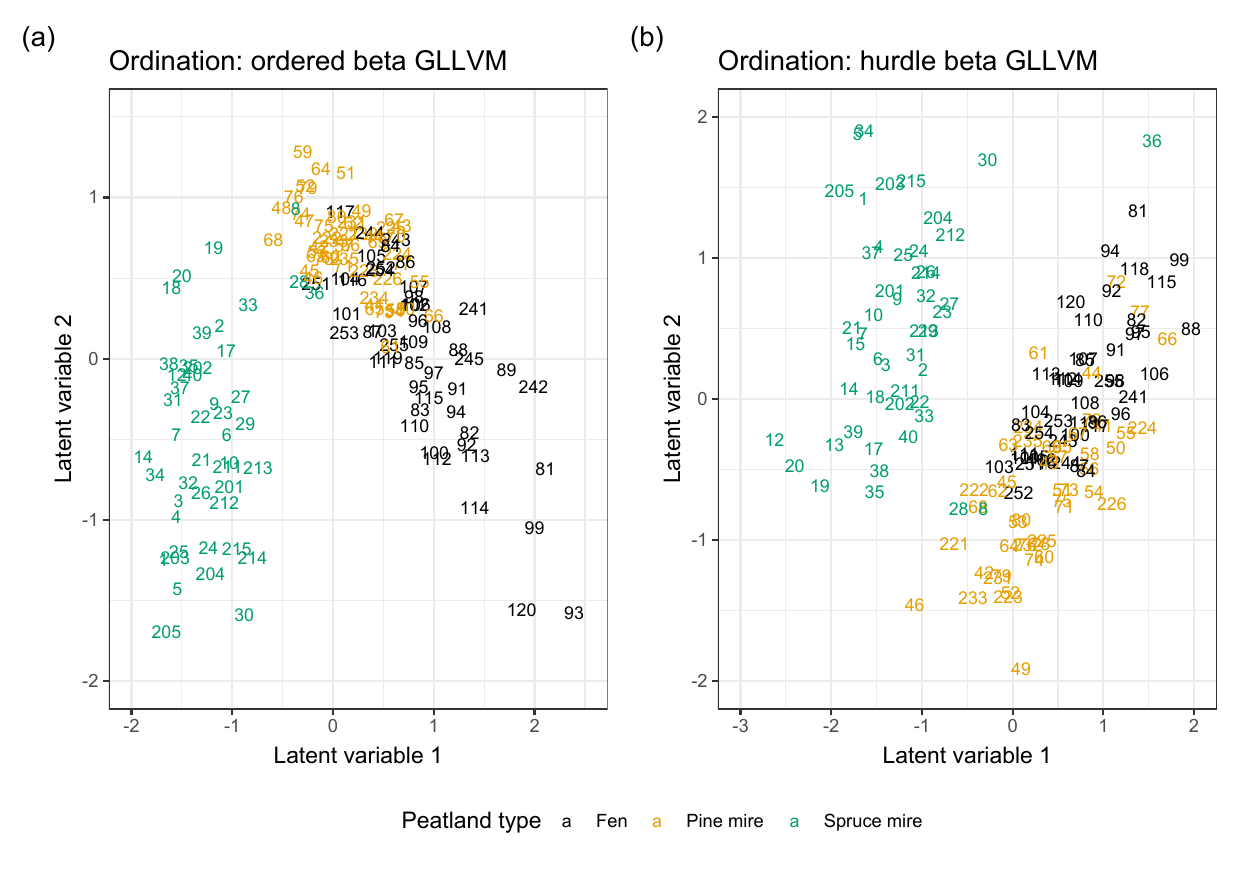}
    \caption{Model-based unconstrained ordination plots based on (a) ordered beta, and (b) hurdle beta GLLVMs fitted to the vascular plant cover dataset. Sites are colored according to their peatland type. These clear clusters in the latent variable scores would dissipate if the peatland type was included in the GLLVM as a covariate.}
    \label{fig:ordination_example}
\end{figure}

For visualizing community composition data, ordination plots display observational units according to their scores on a small set of latent axes, such that units closer to each other in the ordination can be deemed to be more similar in species composition or relative abundance \citep{warton2015so,veen2021}. 
While traditional ordination methods are algorithmic and based on distance measures \citep[e.g., non-metric multidimensional scaling or NMDS,][]{kruskal1964multidimensional, kruskal1964nonmetric},
model-based approaches to ordination using JSDMs have surged in popularity over the past decade due their capacity to (also) incorporate environmental covariates, complex dependence structures, and species' traits and phylogeny information \citep[e.g.,][]{ veen2021,popovic2022fast,van2023concurrent}. 
Fig. \ref{fig:ordination_example} presents an example of model-based unconstrained ordination based on two new forms of GLLVMs we propose for percent cover data, and fitted to a vascular plant dataset from $151$ peatland sites across Finland (see the simulation study later on for further details). The vegetation cover data is extremely sparse with a large proportion of covers recorded to be exactly zero or one. 
When the sites are colored according to peatland type, we see clear clusters forming accordingly -- this showcases the ability of the GLLVMs to account for unobserved characteristics of the data when performing model-based ordination.

The remainder of this article is structured as follows. We begin by introducing GLLVMs as a method for analyzing multivariate percent cover data with exact zeros, and possibly exact ones. Afterward, we briefly review existing approaches for modelling multivariate percent cover data, starting with a GLLVM assuming a beta distribution in combination with a common transformation used for exact zeros and ones \citep{smithson2006better}. We then propose a new three-part hurdle model extension of the beta GLLVM, extending the work of \citet{ospina2012hurdle, Liu2015zoibAR} to the case of multiple correlated responses. We also review the cumulative logit GLLVM for ordinal (cover class), responses, and propose a model inspired by the recent work of  \citet{kubinec_2023} on the ordered beta distribution. 

We perform a series of numerical comparisons between the various JSDMs using both simulated artificial cover data, and by making predictions based on real-world cover data. For the former, we are particularly interested in the different models' ability to accurately recover the latent variables under model misspecification and increasing rate of zeros. Correct retrieval of latent variable scores is an important aspect of a model's performance, due to their part in the resulting model-based ordination.
The comparison also includes the popular algorithmic alternative NMDS for distance-based ordination. For 
predictive comparisons with real-world cover data, we split the datasets into training and test sets, and use various metrics to assess point prediction and classification performance. Special consideration is paid to the effect of the sparsity (or equivalently, recorded prevalence) of the species on predictive performance. We conclude the article with some general remarks and discussion on the results and ideas for future studies.

\section*{GLLVMs for percent cover data} \label{sec:methods}
Cover data in ecology typically comprise records for the proportion of a plot that species, e.g. plants or other sessile organisms, occupy. Denote the coverage of species $j$ in sample $i$ as $Y_{ij}$ for $i=1, \dots, n, j=1,\dots,m$, where $Y_{ij}$ belongs in the closed-interval $[0,1]$. Statistical modelling of proportion data that includes zeros and ones is challenging in general, though a variety of models have been proposed in the literature for univariate responses, i.e., single species. Here we propose a number of extensions for these regression models to the setting of joint species distribution modelling, using GLLVMs, for multivariate percent cover data.

In GLLVMs, we consider regressing the mean of each response $\mu_{ij}=E(Y_{ij})$ against a vector of $d \ll m$ latent variables, $\bo{u}_i=(u_{i1},\dots,u_{id})^\top$, along with the $q$-vector of covariates $\bo x_i=(x_{i1},\dots,x_{iq})^\top$ as follows
\begin{equation} \label{eqn:gllvm}
g(\mu_{ij}) = \eta_{ij} = \alpha_i + \beta_{0j} + \bo x^\top_i \bo\beta_j + \bo u_i^\top\bo\gamma_j,
\end{equation}
where $g(\cdot)$ is a known link function, the vectors $\bo \beta_j=(\beta_{j1},\dots,\beta_{jq})^\top$ and $\bo\gamma_j=(\gamma_{j1},\dots,\gamma_{jd})^\top$ denote species-specific regression coefficients and loadings, respectively, $\beta_{0j}$ denote species-specific intercepts, and $\alpha_i$ denote (optional) row effects.
The latent variables $\bo u_i$ can be considered as ordination scores that capture the correlation across species after accounting for observed covariates $\bo x_i$. These are typically assumed to follow a $d$-dimensional standard normal distribution, $\bm u_i \overset{\mathrm{i.i.d}}{\sim} \mathcal{N}(\bm 0, \textbf{I}_d)$. 

\subsection*{Beta GLLVM}\label{sec:beta}

A popular, starting approach for modelling ecological percent cover data is to use a beta distribution defined for a bounded continuous interval. That is, assume $Y_{ij} \in (0,1)$, meaning it can take any value between but can not exactly be equal to zero or one. Then we assume a beta distribution $Y_{ij} \sim \text{Beta}(\mu_{ij},\phi_j)$ with mean $\mu_{ij}$ and species-specific precision parameter $\phi_j>0$. The probability density function of $Y_{ij}$ is given by 
\begin{equation}
\label{eq:fbeta}
   f_{\text{beta}}(Y_{ij}; \mu_{ij}, \phi_j) = \frac{\Gamma(\phi_j)}{\Gamma(\mu_{ij}\phi_j)\Gamma(\phi_j(1-\mu_{ij}))}Y_{ij}^{\mu_{ij}\phi_j-1}(1-Y_{ij})^{\phi_j(1-\mu_{ij})-1}. 
\end{equation}
For a GLLVM, we can relate $\mu_{ij}$ to the covariates and latent variables using equation \eqref{eqn:gllvm}, where $g(\mu) = \log(\mu/(1-\mu))$ is most commonly set to a logit link function.

As the beta distribution cannot account for zeros (or ones), a common solution is to apply a transformation that shifts the responses slightly away from the bounds. One popular transformation come from \cite{smithson2006better}, who suggested replacing $Y_{ij}$ by $Y_{ij}^* = \big(Y_{ij}(N-1)+0.5\big)/N$ and then modelling $Y_{ij}^*$ using \eqref{eq:fbeta}. Although such a beta GLLVM on transformed responses is simple to fit, and produces credible results when the number of recorded zeros and ones is small, an obvious drawback is that the exact zero and one responses can carry important information which may be lost in the process of the transformation. Put another way, it may be ecologically more reasonable to model the zeros and ones separately rather than through a single continuous distribution. From a statistical perspective, particularly with lots of recorded zeros/ones, the responses are pushed up against the boundary and applying beta GLLVMs on transformed responses does not actually address this issue \citep[see][on the analogous issue of using log transformations for count data]{o2010not,warton2018you}.

\subsection*{Hurdle beta GLLVM}\label{sec:hurdle}

A less heuristic approach to account for recorded zeros and ones in percent cover data is to explicitly model their respective probabilities of arising, and doing so separately from values in between zero and one. For instance, in the broader context of modelling semi-continuous responses where exact zeros can arise \citep[e.g., in biomass][]{Nikuetal:2017}, we can consider a model with two distinct parts:
\begin{align*}
    \mathbb{P}(Y=0) &= \rho, \\[-8pt]
    \mathbb{P}(Y=y | Y > 0) &= (1-\rho) \cdot p(y),
\end{align*}
where $\rho$ controls the probability of a zero occurring and $p(y)$ denotes some generic distribution that can only generate positive values. The above can be extended to more than two parts, and falls under a class of models generally called the \textit{hurdle models} \citep{cragg71hurdle}.

As an aside, note another commonly approach to dealing with excess amount of zeros in ecological modelling is zero-inflated regression \citep[e.g.,][]{martin2005zero,wenger2008estimating}. 
Zero-inflation is most often associated with models for count data, and refers to a sort of mixture model consisting of the base model together with a process that generates additional zeros. Since continuous distributions such as the beta distribution in \eqref{eq:fbeta} by construction can not generate zeros, then in this article we will refrain from using terms associated with zero-inflated regression modelling. 

We propose a new hurdle beta GLLVM that accommodates both recorded zeros and ones in multivariate percent cover data data. The proposed approach builds on the ideas of \citet{ospina2012hurdle,Liu2015zoibAR},
and is comprised of the following three-part beta-distribution-based approach to modelling $Y_{ij}$:
\begin{align}
\label{eq:betahurdle}
    P(Y_{ij};\mu_{ij}, \mu_{ij}^0, \mu_{ij}^1, \phi_j) = \begin{dcases*}
        \mu_{ij}^0, & if $ Y_{ij} = 0 $,\\
        (1-\mu_{ij}^0) \mu_{ij}^1, & if $ Y_{i,j} = 1 $,\\
         (1-\mu_{ij}^0) (1-\mu_{ij}^1) \cdot f_{\text{beta}}(Y_{ij};\mu_{ij},\phi_j), & if $ Y_{ij} \in (0,1). $\\
        \end{dcases*}
\end{align}
Analogous to equation \eqref{eqn:gllvm}, we use a link function to connect the three mean parameters to covariates and latent variables. That is, for the the zero- and one-parts respectively we use $g(\mu_{ij}^0) = \eta_{ij}^0 = \beta_{0j}^0 + \bm x^\top_i\bm \beta_j^0 + \bm u^\top_i \bm \gamma_j^0$, and $g(\mu_{ij}^1) = \eta_{ij}^1 = \beta_{0j}^1 + \bm x^\top_i\bm \beta_j^1 + \bm u^\top_i \bm \gamma_j^1$, and $g(\cdot)$ is set to the logit link function. The quantity $f_{\text{beta}}(Y; \mu, \phi)$ is defined as per equation \eqref{eq:fbeta}.

It is not hard to see the hurdle beta GLLVM works by simultaneously modelling the absences (zeros), full coverage (ones), and percent covers between this. The three linear predictors $\eta^0_{ij}, \eta^1_{ij}$ and $\eta_{ij}$ need not contain the same set of environmental covariates $\bm x_i$ (or even the same latent variables $\bm u_i$). For example, if the data contain a moderate to high rate of zeros but only relatively few ones, then we may choose to use a simple structure for $\eta_{ij}^1$ including the intercept and latent variables only. Also, expert knowledge may inform that a certain covariate is known to have an effect on percent covers of all species present, but does not influence the actual likelihood of presence. Then it might be sensible to leave such as covariate out of $\eta^0_{ij}$ while keeping it in $\eta_{ij}$. This degree of flexibility is a key strength of the hurdle beta GLLVM. For ease of presentation though, and for focusing on the case of producing a single model-based ordination, we have set up the model such that the same $\bm x_i$'s and $\bm u_i$'s occur in all three linear predictors.

\subsection*{Cumulative logit GLLVM}\label{sec:ordinal}

The cumulative logit regression or proportional odds model \citep{mccullagh80} is a popular approach for analyzing ordered categorical, i.e., ordinal, responses. In community ecology, it finds its use in analysing cover class datasets where, instead of cover percentages or counts of specimens, the responses instead comprise labels indicating the class each species is sorted into. Two well-known examples of vegetation cover classification systems are given by \cite{braun1932plant} and \cite{daubenmire1959canopy}.

For each species $j$, assume a classification scheme consisting of levels labeled from $1$ to $(K_j+1)$, ordered from low (or zero) coverage to high (or full) coverage. Then a cumulative logit GLLVM is characterized by the following distribution for the responses:
\begin{equation}
\label{eq:ordinal}
    P\left(Y_{ij}; \eta_{ij}, c_{1}^{(j)}, \dots, c_{K_j}^{(j)}\right) = \begin{dcases*}
        \rho^1_{ij}, & if $Y_{ij}=1$,\\
        \rho^k_{ij}-\rho^{k-1}_{ij}, & if $2\leq Y_{ij} \leq K_j$,\\
        1-\rho^{K_j}_{ij}, & if $Y_{ij}=K_j+1$,
    \end{dcases*}
\end{equation}
where $g(\rho^k_{ij}) = c_k^{(j)} - \eta_{ij}$ with $g(\cdot)$ set to the logit link, $\eta_{ij} = \bm x^\top_i\bm \beta_j + \bm u^\top_i \gamma_j$ where the species-specific intercept is omitted for reasons of parameter identifiability, and $c_1^{(j)} < c_2^{(j)} < \dots < c_{K_j}^{(j)}$ are a set of cutoff parameters specific to the species $j=1,\dots,m$. To ensure the model can be fitted, for every species $j = 1,\ldots,m$ the data must include at least one observation in each of the corresponding $K_j+1$ levels. Note for interpretation, it may make more sense to use a common set of cutoffs for all species i.e. one set of parameters $c_1 < \dots < c_K$ and reintroduce $\beta_{0j}$ e.g., if a common cover class system is used \citep{hui2018boral}. 
On the other hand, while using species-common cutoffs may (also) ease fitting  of the cumulative logit GLLVM, it is less flexibility than allowing species-specific cutoffs as in equation \eqref{eq:ordinal}; it is our recommendation for practitioners to carefully consider the classification systems employed for each species during the data collection process before deciding which particular version of the model to adopt.

\subsection*{Ordered beta GLLVM}\label{sec:ordered}
We propose an extension of the cumulative logit GLLVM, called the ordered beta GLLVM to handle percent cover data that includes exact records of zeros and ones. The proposed approach is based on the idea of an ordered beta distribution \cite{kubinec_2023}, which formulates a conditional cumulative logit process for responses belonging in one of the classes $\big\{\{0\},(0,1),\{1\}\big\}$ i.e., zeros, between zero and one, or ones. Conditional on being in the second class, the percent cover is then represented by a beta distribution. The ordered beta GLLVM thus arguably presents a more ``continuous" data generating process, in contrast to the hurdle beta GLLVM which effectively separates the three types of classes $\big\{\{0\},(0,1),\{1\}\big\}$ into distinct parts. For example, \cite{kubinec_2023} argues that with the hurdle beta model, it is possible for a set of environmental covariates to simultaneously have an increasing effect on the probability of observing zeros and ones, which is usually unrealistic.

We now formulate the ordered beta GLLVM in more detail. For species $j = 1,\ldots,m$, let $z_{ij}$ denote an underlying continuous variable, and define two cutoff parameters $\zeta_{j0} < \zeta_{j1}$ such that $Y_{ij}=0$ occurs when $z_{ij}<\zeta_{j0}$, $Y_{ij}=1$ occurs when $z_{ij}>\zeta_{j1}$, and $Y_{ij} \in (0,1)$ occurs when $\zeta_{j0} < z_{ij} < \zeta_{j1}$. Conditional on $Y_{ij} \in (0,1)$, the response variable follows a beta distribution $Y_{ij}$ as per equation \eqref{eq:fbeta}. 
By assuming $z_{ij}$ follows a logistic distribution, then  marginalizing over $z_{ij}$ we obtain the following distribution for the percent cover responses that characterizes the ordered beta GLLVM, 
\begin{align}
\label{eq:ordbeta}
    P(Y_{ij};\eta_{ij}, \phi_j) = 
    \begin{dcases*}
       \rho^0_{ij}, & if $ Y_{ij} = 0 $,\\
        \left(\rho^1 - \rho^0 \right) \cdot f_{\text{beta}}(Y_{ij}; \mu_{ij}, \phi_j), & if $ Y_{ij} \in (0,1) $,\\
        1-\rho^1_{ij}, & if $ Y_{ij} = 1 $,\\
        \end{dcases*}
\end{align}
where $g(\rho^k_{ij}) = \zeta^{(j)}_k - \eta_{ij}$ and $g(\cdot)$ the logit link as in cumulative logit GLLVM, and $\eta_{ij} = \beta_{0j} + \bm x^\top_i\bm \beta_j + \bm u^\top_i \gamma_j$
The ordered beta GLLVM looks somewhat similar to the case of a cumulative logit GLLVM in equation \eqref{eq:ordinal} with three classes, except the middle class is coupled with a standard beta regression model as in \eqref{eq:fbeta}. In doing so, the ordered beta GLLVM greatly reduces the total amount of parameters to be estimated compared with the hurdle beta GLLVM in \eqref{eq:betahurdle}. This may prove advantageous in situations where the latter tends to overfit or there is not enough information in the multivariate percent cover data to adequately model the probability of zeros and ones separately. Here, only one linear predictor is needed to model all parts of the data, and connects all three possible ``states" of the response (either it is recorded as exactly zero, or between zero and one, or exactly one). On the other hand, 
in situations where the data \emph{does} carry enough information about the distinct the ecological processes generating the zeros, ones or percent cover continuous responses, the hurdle beta GLLVM is expected to perform similar to or better than its ordered beta counterpart.

\subsection*{Model fitting}\label{sec:fitting}

When working with non-continuous responses and a non-identity link function/s, GLLVMs can be estimated using approximate likelihood-based methods, with the approximation arising since the marginal log-likelihood lacks a tractable form. One class of approaches that has garnered a lot of attention recently is variational approximations for GLLVMs, which have shown to be an attractive choice over alternatives such as Laplace approximation or quadrature rules \citep{korhonen2023}; see also the recent quasi-likelihood approach of \citet{kidzinski2022generalized}. For the particular models studied in this article, we use 
the class of extended variational approximations (EVA) described in \cite{korhonen2023}, which allows efficient approximate likelihood-based fitting and inference even when tractable, closed-form expressions can not be immediately obtained. We also coded every model in a similar manner, utilizing \texttt{TMB} for its fast implementation of automatic differentiation, so as to ensure any comparisons we make are not confounded by differences in estimation and inferential methods.

\section*{Numerical study}

\subsection*{Simulation design}

With a focus on model-based ordination, we used two simulation setups to compare the Procrustes error \citep{bartholomew2011} between the predicted and true latent variables under increasing degrees of sparsity for multivariate percent cover data. We compared latent variables obtained from three different beta-distribution based GLLVMs (shifted, ordered, hurdle), the cumulative logit model GLLVM on ordinally classified responses, and a Bernoulli logit GLLVM on presence-absence transformed responses (i.e., the percent cover was converted to zeros and ones). We also included non-metric multidimensional scaling (NMDS) with either the Bray-Curtis or Jaccard dissimilarity measures, as an algorithmic distance-based alternative to ordination.

The first and second simulation setups used the ordered beta GLLVM and the hurdle beta GLLVM, respectively, as the true data generating processes. In both of these cases, we simulated $1500$ multivariate percent cover datasets of $[0,1]$-responses with $n=180$ (units) and $m=240$ (species). We considered the mean proportion of zero observations across the $m$ species to be varying as $p=10\%, 20\%, \dots, 90\%$, while the proportion of ones was kept constant at $5\%$. When fitting the cumulative logit GLLVM, we assumed common cutoff parameters for all species and converted the simulated percent cover responses to class numbers in accordance with the Daubenmire system.
For simplicity, both simulation setups featured no predictor variables, with the species-specific intercepts $\beta_{0j}$ drawn from a uniform distribution $\mathcal{U}(-1,1)$, the elements of the loading vectors $\bm \gamma_j$ drawn independently from $\mathcal{U}(-2,2)$, and the latent variables $\bm u_i$ drawn from $\mathcal{N}(\bm 0, \mathbf{I}_2)$. When simulating from the hurdle beta GLLVM, the additional loadings $\bm \gamma^0_j$ and $\bm \gamma^1_j$ were also drawn from $\mathcal{U}(-2,2)$, while the cutoff parameters $\zeta^{(j)}_{k}$ and additional intercepts $(\beta_{0j}^0,\beta_{0j}^1)$ in the true ordered and hurdle beta GLLVMs respectively were chosen to best fulfil the desired proportions of zeros and ones as discussed above.

\begin{figure}[htb]
    \centering     \includegraphics[width=\linewidth]{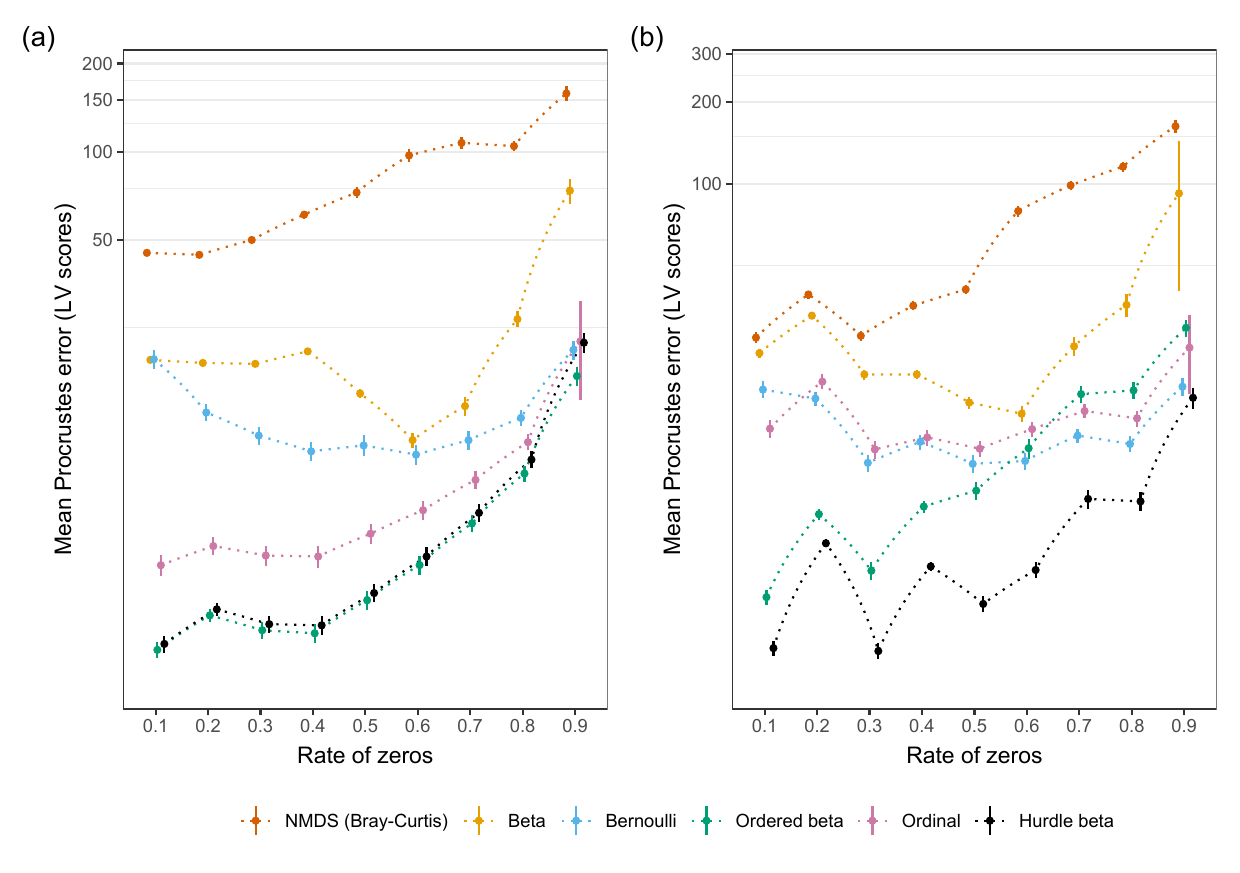}
    \caption{Means and standard deviations of the Procrustes errors between the predicted and the true latent variable scores, where multivariate percent cover data were generated using the: (a) ordered beta GLLVM, and (b) hurdle beta GLLVM. The most extreme 5\% of the values were discarded. The points are slightly jittered to avoid visual overlap.}
    \label{fig:simres}
\end{figure}

\subsection*{Results}

Results for the Procrustes error are shown in Fig. \ref{fig:simres}, noting that NMDS with Jaccard dissimilary metric was omitted due to it performing almost identically to the Bray-Curtis dissimilarity. 
Overall, NMDS consistently performed the worse at recovering the true latent variables values. Of the model-based approaches, 
the simple beta GLLVM with transformed responses struggled the most especially when the mean proportion of zeros was quite high and the data generating model was hurdle beta GLLVM (Fig. \ref{fig:simres}{b}). Unsurprisingly, the performance of the Bernoulli model shows improvement as the simulated responses start to resemble presence-absence data more, i.e., as the sparsity level increases. Generally, the hurdle beta GLLVM performed best across both scenarios, which was to be expected given it is more flexible than the ordered beta GLLVM and was the true model in second simulation setup. Finally, the cumulative logit GLLVM performed only slightly worse than its ordered and hurdle beta counterparts across the board in the first setup (Fig. \ref{fig:simres}{a}), but was noticeably worse than these two models in the second scenario when the data was dense i.e., the mean proportion of zero responses was low (Fig. \ref{fig:simres}{b}). Such a result however can be attributed more to the behaviour of the ordered and hurdle beta GLLVMs themselves: when the multivariate percent cover data was relatively dense, these two models outperform the cumulative logit GLLVM since converting such data to cover classes results in a non-negligible loss of information in the responses. However as the data became more sparse (and discrete) the performance of the ordered and hurdle beta GLLVMs starts to deteriorate while cumulative logit GLLVM continues to perform similarly (since in this case converting to cover classes does not lose as much information).

\section*{Application and comparison of predictive performances}\label{sec:predictions}

\subsection*{Data and setup}
\label{sec:pred_settings}

We compared the predictive capabilities of the three beta-distribution-based GLLVMs on four real multivariate percent cover data sets. The first two datsets originate from the Santa Barbara Coastal Long Term Ecological Research \citep[SBC LTER,][]{SBC_LTER} site, where cover percentages of more than 150 sessile invertebrates and macroalgae taxa were recorded between years 2000 and 2020, along $40$m $\times$ $2$m permanent transects at 11 kelp forest sites across coastal Southern California. The number of transects differed among sites, varying between two and eight with a total of 44, while the two datasets were defined separately for algae and invertebrates. About half of the taxa corresponding to extremely rare species were removed prior to analysis. 
When fitting the various GLLVMs, we set the latent variables to be at the transect level, leading to a total of $n = 44$ two-dimensional latent variables scores to estimate. Each model was fitted to data recorded from 2000-2017, with years 2018-2020 held out to assess predictive performance. The two datasets also included two key environmental covariates: rockiness of the seabed, and the number of stripes of giant kelp. These were used together with year as covariates in $\bm x_i$. 

The second pair of datasets come from a Finnish longitudinal study on effects of peatland restoration, comprising over 250 species of vascular plants and mosses collected on 151 sites across Finland between years 2006 and 2022 \citep{eloetal:2016}. Within each site, the percent cover of each species was measured on 10 placed plots of size $1$m$^2$. The sites varied on a number of key environmental factors: type (fen, pine mire and spruce mire), treatment level (drained, pristine and restored), and productivity level (high vs. low). Similar to the two SBC LTER datasets, the two datasets were defined separately for vascular plant and moss species, and in both we subset to only species for which there existed at least one observation per each environmental factor level. The 
latent variables were estimated at the site level, resulting in a total of $n =151$ two-dimensional ordination scores to estimate. We estimated GLLVMs using data from 2006-2021, and held out data from the final year (2022) to assess predictive performance.

We assessed performances of the fitted GLLVMs using four metrics: mean absolute error of prediction (MAEP) and root mean square error of prediction (RMSE), which were calculated for the beta, hurdle beta, and ordered GLLVMs and for each species individually, and area under the receiver operating characteristic curve (AUC) and Tjur's pseudo-R$^2$ \citep{tjurr2} for classifying presences and absences. The latter two measures assess capacity to discriminate between zeros and ones, and were calculated for binary, ordinal, ordered beta and hurdle beta GLLVMs.
With the ordinal GLLVM, we assumed species-common cutoff parameters and used seven classes with the following bounds: $\{0\},(0,0.02], (0.02,0.05],(0.05,0.25],(0.25, 0.5], (0.5, 0.7]$, and $(0.7,1]$. We note the selection of classification scale is expected to have an impact on performance of the ordinal model, and our choice here is based on exploratory data analysis. Future research could examine how sensitive prediction, interpretation, and statistical inference for ordinal models are in general to the scale of classification.

As all of the data sets either entirely lacked, or had very small amount, of records exactly equal to one (full cover of a measurement area), then for the hurdle beta GLLVM in equation \eqref{eq:betahurdle} we only included the part for modelling zeros and values strictly between zero and one. Furthermore, for the ordered beta GLLVM in \eqref{eq:ordbeta}, we used species-common upper cutoff parameters. 

\subsection*{Results}
\label{sec:pred_results}

\begin{figure}[ht]
    \centering
    \includegraphics[width=\linewidth]{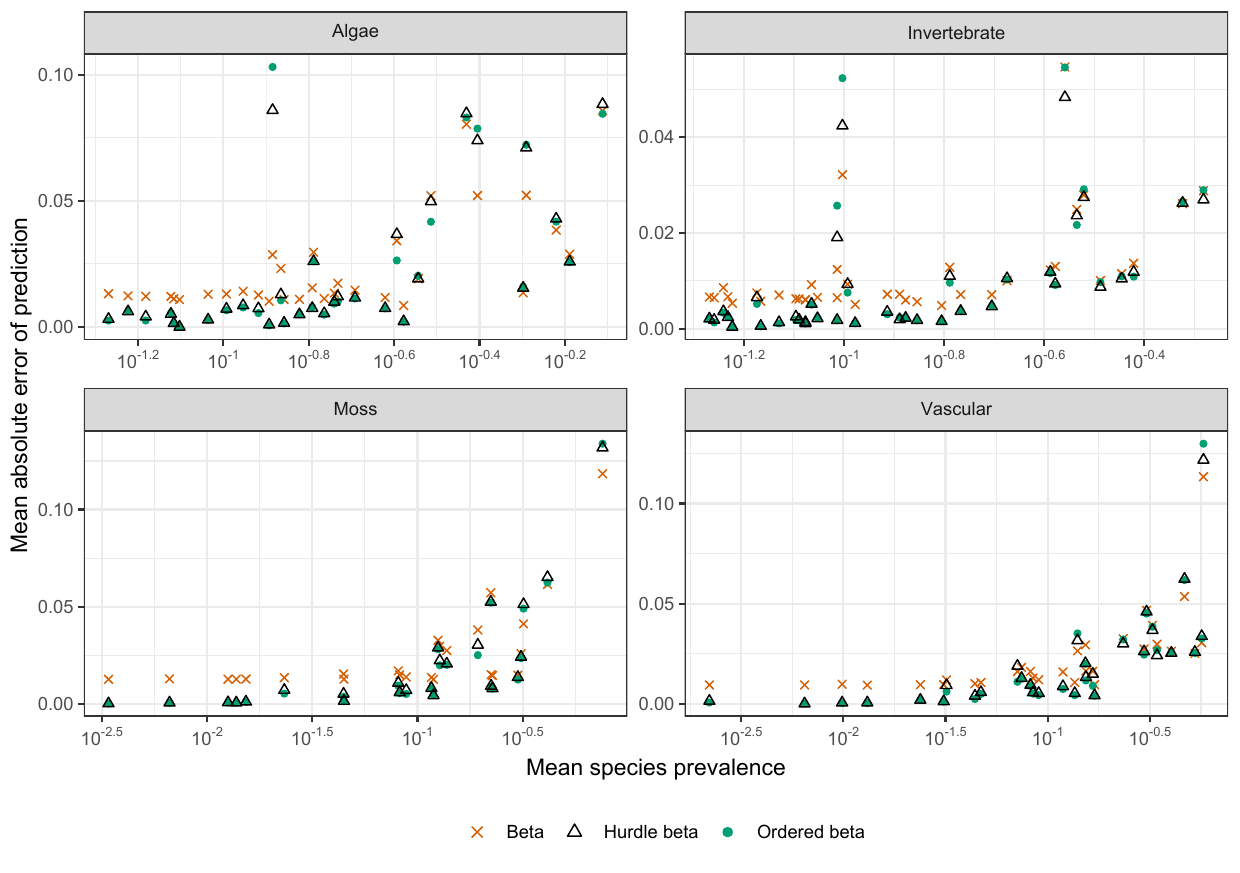}
    \caption{Mean absolute error of prediction (MAEP) as a function of mean species prevalence for beta, hurdle beta, and ordered beta GLLVMs, across the four real multivariate percent cover datasets.}
    \label{fig:prev_maep_log10}
\end{figure}

\begin{figure}[ht]
    \centering
    \includegraphics[width=1\linewidth]{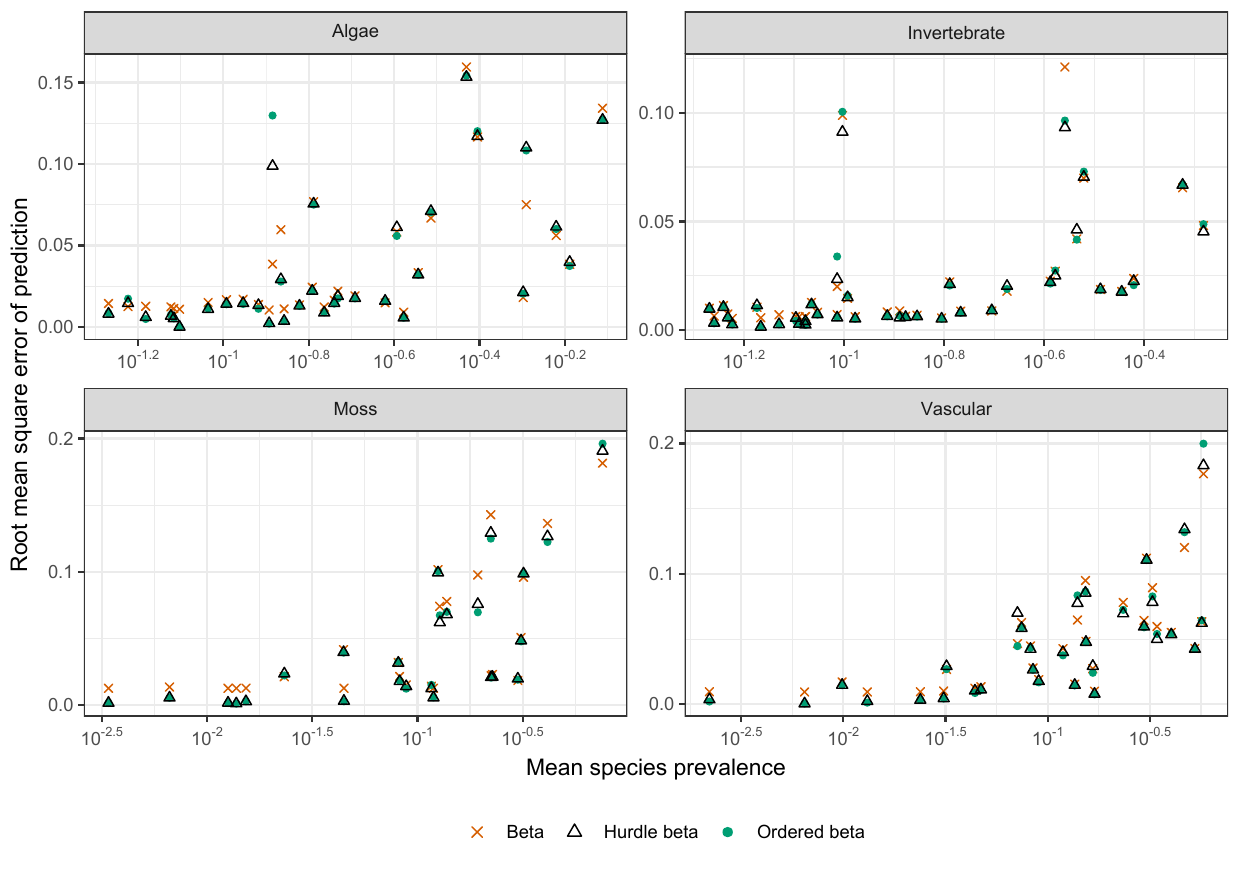}
    \caption{Root mean square error of prediction (RMSE) as a function of mean species prevalence for beta, hurdle beta, and ordered beta GLLVMs, across the four real multivariate percent cover datasets.}
    \label{fig:prev_rmse_log10}
\end{figure}

Figs \ref{fig:prev_maep_log10}-\ref{fig:prev_rmse_log10} plot the MAEP and RMSE, respectively, as a function of total species prevalence, $p$,  across each of the four datasets. Across both figures, the beta GLLVM tended to perform worse than the two zero-accommodating approaches when predicting for the rarer species, which 
is to be expected given the capacity of the hurdle and ordered betas GLLVMs to systemically handle the recorded zeros. The differences between these three models diminished when predicting for the more commonly occurring species. Note the actual metrics are small in magnitude for the rarer species overall; this is generally not surprising given the more discrete nature responses in such cases. 
Between the hurdle and ordered beta GLLVMs, there were no noticeable differences in performance.

\begin{figure}[htb]
    \centering     \includegraphics[width=1\linewidth]{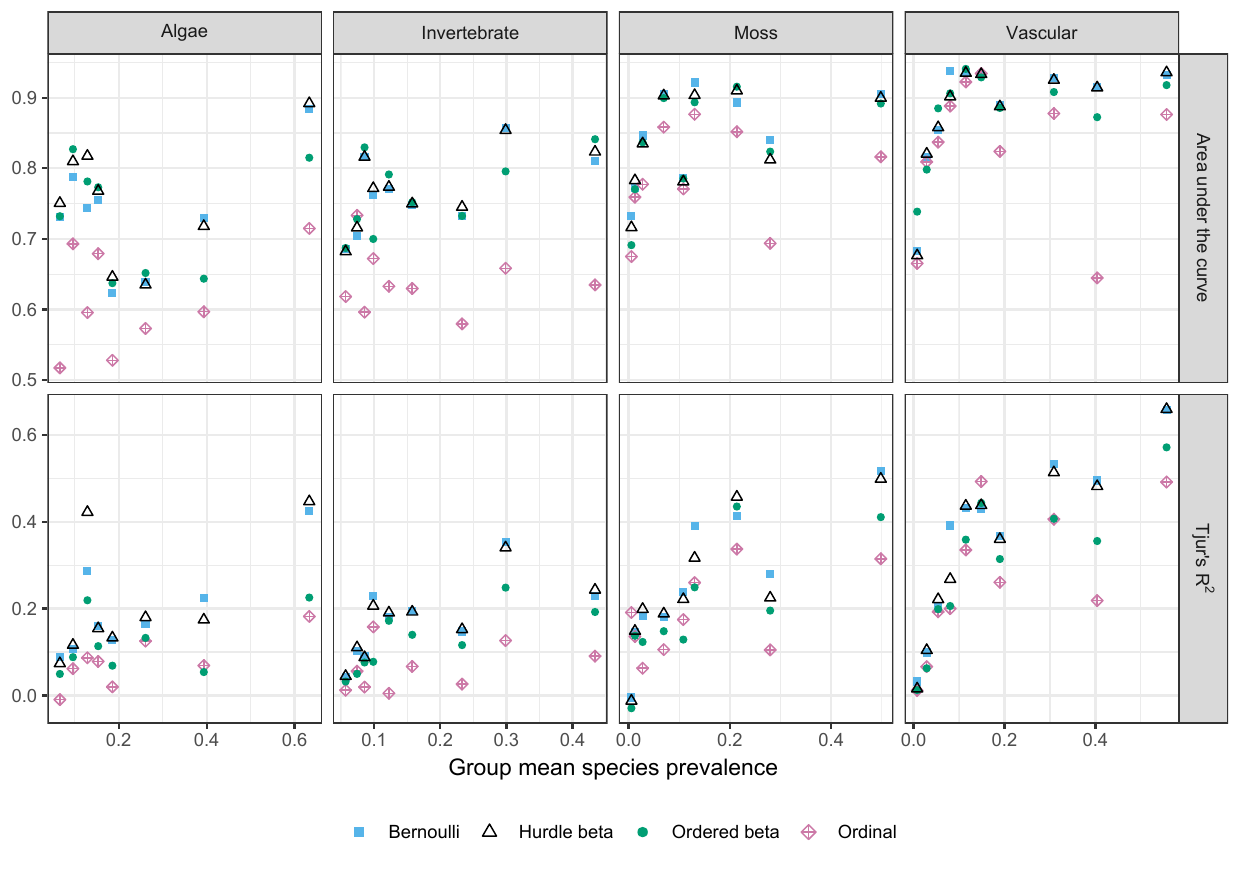}
    \caption{AUC (top row) and Tjur's $\text{R}^2$ (bottom row) as a function of recorded group mean prevalence for the four real multivariate percent cover datasets. Recorded group mean prevalence was obtained by clustering species based on their recorded prevalences in the complete dataset into a small number of groups, and then calculating the mean prevalence of each group. The $y$-axes presents the corresponding metric for each group.}
    \label{fig:betapred_prev}
\end{figure}

Fig. \ref{fig:betapred_prev} presents the values of AUC and Tjur R$^2$ across the four datasets when plotted against the recorded group mean prevalence. For each dataset, the recorded group mean prevalence was obtained by computing the proportion of non-zero observations in the complete dataset for each species, clustering species based on these recorded prevalences into a small number of groups, and then calculating the mean prevalence of each group. The values of AUC and Tjur R$^2$ were then calculated correspondingly for each group. Note since both metrics are determined based on estimated probabilities of presence, then only models capable of producing these estimates were included for comparison, i.e., the beta GLLVM is omitted. 
Overall, the hurdle beta GLLVM model followed the Bernoulli logit model closely in its ability to best discriminate between zero and non-zero responses across all levels of prevalence. The ordered beta GLLVM performed similarly to the hurdle and Bernoulli GLLVMs in terms of AUC, although its performance was closer to the ordinal/cumulative logit GLLVM, which performed worst overall, when it came to Tjur R$^2$. 


\section*{Discussion}
\label{sec:discussion}

In this paper, we compared different joint species distribution modelling approaches for multivariate percent cover data. Such data are encountered, for example, in studies where percent cover of several plants or other sessile organisms is measured across multiple sites, as per the four real datasets we investigated. The typical ecological questions arising from such studies include whether sites exhibit similar species composition, how environmental factors influence community composition, 
and whether we can predict vegetation or macroalgal cover at new sites and/or over time. This article focused on comparing methods in terms of community composition analysis through model-based ordination, and in making out-of-sample predictions. We explored and extended several of GLLVMs to handle percent cover data with exact zeros and ones, most notably 
a three-part hurdle beta GLLVM, a cumulative logit GLLVM for cover class responses, and an ordered beta GLLVM. As ecological percent cover data are often sparse with a large number of recorded zeros along with potentially a number of full coverages (recorded ones), it was of primary interest to study how methods performed when observed species prevalence decreased.  
When comparing ordinations using simulation studies, we found the hurdle beta GLLVM exhibited the best overall performance, while the classical beta GLLVM on transformed responses performed poorly. Both the hurdle beta and the ordered beta GLLVMs performed relatively well when it came to prediction and classification, and better than the cumulative logit and the classical beta GLLVMs. We anticipate these results would be alike were similar comparisons done on some other families of JSDMs, e.g., copula models \citep{popovic2022fast} or basis-function models \citep{hui23cbfm}. Implementations of all of the latent variable models compared in this article are readily available in the \texttt{gllvm} \textsf{R}-package \citep{gllvmpackage}

One area of future research concerns extending the GLLVM framework for alternative types of cover data, namely data collected using the pin-point method or for compositional data increasingly seen in community ecology.
For such data, there exists several competing frameworks, e.g., classic log-ratio analysis \citep{aitchison1982statistical}, regression models based on Dirichlet or Dirichlet-multinomial distributions \citep{douma2019analysing, damgaard2020model, kettunen23jsdm}, and distributions directly on the composition itself \citep[e.g.,][]{scealy2014fitting,scealy2023score}. 
With most of these approaches, incorporating structural zeros comes with challenges; the standard log-ratio transformations of Aitchison geometry are not defined for zero observations, and the so-called $\alpha$-transformation \citep{clarotto2022alpha} introduces an additional parameter which may be difficult to optimize for in sparse and high-dimensional settings often encountered in community ecology. Similarly, the standard Dirichlet or Dirichlet-multinomial distributions are also incapable of handling structural zeros, and implementing e.g., (multinomial-)Dirichlet hurdle model in the fashion of \eqref{eq:betahurdle} in likelihood-based setting requires may not be straightforward; see also \cite{tang2018zigdm} for a Bayesian treatment of such models.

Extending the ordered beta and hurdle beta GLLVMs to handle spatially or spatio-temporally dependent latent variables could also prove fruitful; presumably, ecosystems closer to each other geographically are expected to also be more alike in the community structures. 
Similarly, 
many multivariate percent cover datasets originate from longitudinal studies, including both the SBC LTER \citep{SBC_LTER} and the Finnish peatland \citep{eloetal:2016} datasets considered in this article. Estimating parameters for general spatio-temporal covariance structures is typically very challenging in high-dimensional settings, and such GLLVMs would most likely require employing additional approximation techniques, e.g., nearest neighbour Gaussian processes \citep[e.g.,][]{Tikhonovetal:2020spat,lawler2023starve} or some form a basis-function/fixed-rank kriging approach \citep[e.g., see the recent work of][]{hui23cbfm,ning2023double}.

\section*{Acknowledgements}
PK was funded by the Wihuri Foundation (00220161), and PK, JN and ST were funded by the Kone foundation (201903741). ST was funded by the Research Council of Finland (453691) and the HiTEc COST Action (CA21163). FKCH was funded by an Australian Research Council Discovery Project (DP230101908). We thank Merja Elo  at University of Finland and Santtu Kareksela at Metsähallitus Parks \& Wildlife Finland for providing us the Finnish peatland dataset. 

\section*{Author Contributions}
 All authors conceived the idea for the manuscript, PK and JN implemented the methodology, and performed all performance studies. PK led the writing of the manuscript, while all authors contributed to the drafts and gave final approval for publication.

\section*{Data Availability}
The kelp forest community data \citep{SBC_LTER} are available in SBC LTER database \url{https://doi.org/10.6073/pasta/0af1a5b0d9dde5b4e5915c0012ccf99c}. The plant and moss community data from  Finnish peatlands are available from the author's of \cite{eloetal:2016} upon request. The Bernoulli logit and beta GLLVMs models are available as part of the \textsf{R}-package \texttt{gllvm}, while implementations for the hurdle beta, ordered beta and cumulative logit GLLVMs are available at \url{https://github.com/JenniNiku/gllvm}. NMDS is implemented in the \texttt{vegan} \textsf{R}-package \citep{vegan13}.

\FloatBarrier

\bibliographystyle{apalike} 
\bibliography{ref}

\end{document}